\documentclass[aps,prb,twocolumn,floatfix,superscriptaddress,showpacs,Superscript citations]{revtex4-1}

\usepackage{appendix}
\usepackage{graphicx}
\usepackage{amsfonts}
\usepackage{amsmath}
\usepackage{amssymb}
\usepackage{bm}
\usepackage{color}
\usepackage[hypertex]{hyperref}%
\providecommand{\U}[1]{\protect\rule{.1in}{.1in}}

\begin{document}

\title{Josephson current through a ferromagnetic bilayer: Beyond the quasiclassical approximation}
\author{Hao Meng}
\affiliation{School of Physics and Telecommunication Engineering, Shaanxi University of Technology, Hanzhong 723001, China}
\affiliation{University Bordeaux, LOMA UMR-CNRS 5798, F-33405 Talence Cedex, France}
\author{Yajie Ren}
\affiliation{School of Physics and Telecommunication Engineering, Shaanxi University of Technology, Hanzhong 723001, China}
\author{Javier E. Villegas}
\affiliation{Unit\'{e} Mixte de Physique CNRS/Thales, Universit\'{e} Paris-Sud, Universit\'{e} Paris Saclay, 1 Avenue A. Fresnel, 91767 Palaiseau, France}
\author{A. I. Buzdin}
\email{alexandre.bouzdine@u-bordeaux.fr}
\affiliation{University Bordeaux, LOMA UMR-CNRS 5798, F-33405 Talence Cedex, France}
\affiliation{Sechenov First Moscow State Medical University, Moscow, 119991, Russia}

   \date{\today}

   \begin{abstract}
    Based on the Bogoliubov-de Gennes equations, we provide an exact numerical solution for the critical current of Josephson junctions with a composite ferromagnetic bilayer. We demonstrate that for the antiparallel orientation of the magnetic moments of the bilayer, the presence of a potential barrier at the bilayer interface results in large oscillations of the critical current as a function of ferromagnet thickness and/or exchange field. Because of this, and remarkably, in the range of small exchange field and thicknesses, the magnetism leads to the increase of the critical current. This effect is well pronounced at low temperature but disappears near $T_c$. If the potential barrier is replaced by a spin-active barrier at the bilayer interface the conventional 0-$\pi$ transition, similar to the case of an uniform ferromagnetic Josephson junction, is observed. Strikingly, for a parallel orientation of the magnetic moments of the bilayer, the presence of the spin-active barrier restores the anomalous behavior---potential barrier in the antiparallel case. These behaviors result from the resonant tunneling of Cooper pairs across the composite barrier---an effect related to the spin-dependent Fermi vector in the presence of the ferromagnets' exchange field.
   \end{abstract}

   \maketitle

   \section{Introduction}

    In recent years the superconductor (S)-ferromagnet (F) systems attracted a lot of attention due to the possibility to fabricate the new devices based on the superconducting spintronics~\cite{Golubov,Buzdin,Bergeret,Linder,Eschrig}. The properties of different S/F systems may be rather well qualitatively understood in the framework of quasiclassical Eilenberger~\cite{Eilenberger} and Usadel~\cite{Usadel} approaches. However, the applicability of these methods assumes that the exchange field $h$ in the ferromagnet should be much smaller than the Fermi energy $h\ll{E_{F}}$ and the use of the Usadel equations implies even more restrictive conditions $h\tau\ll1$, where $\tau$ is the electrons scattering time. These circumstances lead to the fact that some subtle qualitative effects may be missed by the quasiclassical approach, see, for example~\cite{Reeg,Silaev,HaoBuzdin}. Moreover, a lot of experimental activities with the S/F heterostructures deal with the strong ferromagnets (or even half-metals~\cite{CVisani,CVFC,MJWA}) for which the quasiclassical approximation cannot provide an adequate quantitative description.

    The alternative approach for the analysis of proximity effects in strong ferromagnets is the use of the microscopical approach on the basis of the Bogoliubov-de Gennes (BdG) equations~\cite{PGdeGennes}. The exact numerical solutions of these equations may provide additional information to the quasiclassical approach and this method was used in~\cite{CitE,CitC,CitB,CitA,CitD,Hman2,Hman3,Hman4,Hman5,Hman6} and references cited therein. Recently the interesting experimental results were obtained for the Josephson junctions containing a ferromagnetic spin valve~\cite{CBell,JWAR, BBa,MAEQ,BBWHR,ECGing,BMNied}. Taking in mind these experiments in the present work we study the SFS junctions with composite F layer consisting of two parallel or antiparallel ferromagnetic layers separated by either a potential or a spin-active barrier.

    Note that previously the Josephson junctions with ferromagnetic bilayers were studied theoretically by different methods~\cite{CitE,CitC,CitB,CitA,CitD,CitF,CitG,FSBerg,VNKrivo,Elena,AAGolu,YaMB,BCrou}. However, most of the theoretical analysis was made in the framework of the quasiclassical approach, while in the present work we discuss some effects which cannot be found by this approach and has not been discussed before. We have calculated the critical current of the Josephson junctions with a composite (spin-valve) F$_1$F$_2$ interlayer and studied the role of the potential and spin-active barrier at F$_1$/F$_2$ interface. The obtained results show an anomalous behavior of the critical current as a function of the exchange field and/or F layer thickness which is very sensitive to the type of barrier at the F$_1$/F$_2$ interface.

   \section{Model and formula}

\begin{figure}[ptb]
     \centering
     \includegraphics[width=3.3in]{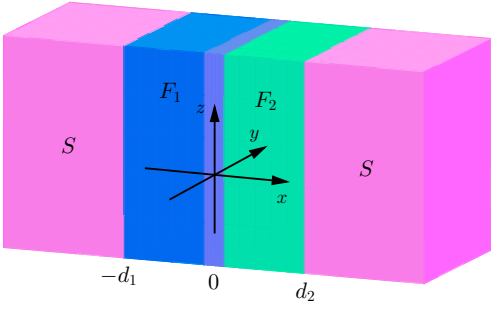}
     \caption{The sketch of SF$_{1}$F$_{2}$S Josephson junction with a potential or spin-active barrier at F$_{1}$/F$_{2}$ interface. The lengths of F$_{1}$ and F$_{2}$ are denoted by $d_{1}$ and $d_{2}$, respectively.}
    \label{Fig1}
    \end{figure}

   The considered SF$_{1}$F$_{2}$S Josephson junction with a central potential or spin-active barrier is shown schematically in Fig.~\ref{Fig1}. The $x$ axis is chosen to be perpendicular to the layer interfaces with the origin located at the central F$_{1}$/F$_{2}$ interface. The BCS mean-field effective Hamiltonian is~\cite{Buzdin,PGdeGennes}
   \begin{align}
     H_{\rm {eff}}=&{\displaystyle\sum\limits_{\alpha,\beta}}\int{d}\mathbf{r}\left\{
     \hat{\psi}_{\alpha}^{\dagger}(\mathbf{\mathbf{r}})\left[H_{e}-(h_{z}
     \hat{\sigma}_{z})_{\alpha\alpha}\right]\hat{\psi}_{\alpha}(\mathbf{r})\right.
     \nonumber\\
     &+\frac{1}{2}\left[(i\hat{\sigma}_{y})_{\alpha\beta}\Delta(\mathbf{r}
     )\hat{\psi}_{\alpha}^{\dagger}(\mathbf{r})\hat{\psi}_{\beta}^{\dagger}(\mathbf{r})+{\rm H.c.}\right] \nonumber\\
     &\left.+\hat{\psi}_{\alpha}^{\dagger}(\mathbf{r})\left(U\hat{\sigma}_{0}-\vec{\rho}\cdot\vec{\sigma}\right) _{\alpha\beta}\hat{\psi}_{\beta}(\mathbf{r})\right\} ,\label{HEFF}
   \end{align}
   where $H_{e}=-\frac{\hbar^{2}\nabla^{2}}{2m}-E_{F}$, and $\hat{\psi}_{\alpha}^{\dagger}(\mathbf{r})$ and $\hat{\psi}_{\alpha}(\mathbf{r})$ represent creation and annihilation operators with spin $\alpha$. $\sigma_{0}$ denotes a $2\times{2}$ unit matrix, and $\vec{\sigma}=\left(\hat{\sigma}_{x},\hat{\sigma}_{y},\hat{\sigma}_{z}\right)$ is the vector of Pauli matrices. Here $m$ denotes the effective mass of the quasiparticles in both the superconductors and the ferromagnets and $E_{F}$ is the Fermi energy. We assume equal Fermi energies in the different regions of the junction. The superconducting gap is supposed to be constant in the superconducting leads and absent inside the ferromagnetic region:
   \begin{equation}
      \Delta(\mathbf{r})=\left\{
      \begin{array}
         [c]{lcl}
         {\Delta}e^{i\phi/2}, &  & x<-d_{1}\\
         0, &  & -d_{1}<x<d_{2}\\
         {\Delta}e^{-i\phi/2}, &  & x>d_{2},
      \end{array}
      \right.
    \end{equation}
     where $\Delta$ is the magnitude of the gap, and $\phi$ is the phase difference between the two superconducting leads. This approximation is justified when, for example, the width of the superconducting layers is much larger than the width of F layers. We model the central F$_{1}$/F$_{2}$ interface by a $\delta$ function potential barrier which consists of a spin-independent part $U=V_{0}\delta(x)$ and a spin-active part $\vec{\rho}=(\rho_{x}, \rho_{y}, \rho_{z})\delta(x)$. The exchange field in two ferromagnetic layers is parallel or antiparallel to the $z$ axis. It has the form
     \begin{equation}
        {h}_{z}=\left\{
        \begin{array}
           [c]{lcl}
           h_{1}\hat{z}, &  & -d_{1}<x<0\\
           \pm{h_{2}}\hat{z}, &  & 0<x<d_{2},
        \end{array}
        \right.
     \end{equation}
     where $\hat{z}$ is the unit vector along the $\emph{z}$ axis.

    To diagonalize the effective Hamiltonian, we use the Bogoliubov transformation $\hat{\psi}_{\alpha}(\mathbf{r})=\sum_{n}[u_{n\alpha}(\mathbf{r})\hat{\gamma}_{n}+v_{n\alpha}^{\ast}(\mathbf{r})\hat{\gamma}_{n}^{\dag}]$ and take into account the anticommutation relations of the quasiparticle annihilation operator $\hat{\gamma}_{n}$ and creation operator $\hat{\gamma}_{n}^{\dag}$. Using the presentation $u_{n\alpha}(\mathbf{r})=u_{p}^{\alpha}e^{ipx}$, $v_{n\alpha}(\mathbf{r})=v_{p}^{\alpha}e^{ipx}$, the resulting Bogoliubov--de Gennes (BdG) equations can be expressed as~\cite{PGdeGennes}
    \begin{equation}%
        \begin{pmatrix}
           \hat{H}_{0}+\hat{V}\delta(x) & i\hat{\sigma}_{y}\Delta(x)\\
           -i\hat{\sigma}_{y}\Delta^{\ast}(x) & -\hat{H}_{0}-\hat{V}^{\ast}\delta(x)
        \end{pmatrix}
        \begin{pmatrix}
           \hat{u}(x)\\
           \hat{v}(x)
        \end{pmatrix}
        =\epsilon
        \begin{pmatrix}
           \hat{u}(x)\\
           \hat{v}(x)
        \end{pmatrix},\label{BdG}
    \end{equation}
    where
    \[
    \hat{H}_{0}=
    \begin{pmatrix}
        \xi_{p}-{h}_{z} & 0\\
        0 & \xi_{p}+{h}_{z}%
    \end{pmatrix},
    \]
    and
    \[
    \hat{V}=
    \begin{pmatrix}
        V_{0}-{\rho}_{z} & -(\rho_{x}-i\rho_{y})\\
        -(\rho_{x}+i\rho_{y}) & V_{0}+{\rho}_{z}%
    \end{pmatrix}.
    \]
    Here $\xi_{p}=\frac{\hbar^{2}p^{2}}{2m}-E_{F}$, and $\hat{u}(x)=[u_{p}^{\uparrow}(x),u_{p}^{\downarrow}(x)]^{T}$ and $\hat{v}(x)=[v_{p}^{\uparrow}(x),v_{p}^{\downarrow}(x)]^{T}$ are quasiparticle and quasihole wave functions, respectively.

    The BdG equation (\ref{BdG}) can be solved for each superconducting electrode and each ferromagnetic layer, respectively. For a given energy $\epsilon$ in the superconducting gap, we find the following plane-wave solutions in the left superconducting electrode:
    \begin{align}
       \psi_{L}^{S}(x)&=C_{1}\hat{\zeta}_{1}e^{-ik_{S}^{+}x}+C_{2}\hat{\zeta}
       _{2}e^{ik_{S}^{-}x}\label{functionSL}\\
       &+C_{3}\hat{\zeta}_{3}e^{-ik_{S}^{+}x}+C_{2}\hat{\zeta}_{4}e^{ik_{S}^{-}x},\nonumber
    \end{align}
    where $k_{S}^{\pm}=k_{F}\sqrt{1\pm{i}\sqrt{\Delta^{2}-\epsilon^{2}}/E_{F}-(k_{\parallel}/k_{F})^{2}}$ are the longitudinal components of the wave vectors for quasiparticles in both superconductors. $\hat{\zeta}_{1}=[1,0,0,R_{1}e^{-i\phi/2}]^{T}$, $\hat{\zeta}_{2}=[1,0,0,R_{2}e^{-i\phi/2}]^{T}$, $\hat{\zeta}_{3}=[0,1,-R_{1}e^{-i\phi/2},0]^{T}$, and $\hat{\zeta}_{4}=[0,1,-R_{2}e^{-i\phi/2},0]^{T}$ are the four basis wave functions of the left superconductor, in which $R_{1(2)}=(\epsilon\mp{i}\sqrt{\Delta^{2}-\epsilon^{2}})/\Delta$. The corresponding wave function in the right superconducting electrode can be described by
    \begin{align}
       \psi_{R}^{S}(x)&=D_{1}\hat{\eta}_{1}e^{ik_{S}^{+}x}+D_{2}\hat{\eta}_{2}e^{-ik_{S}^{-}x} \label{functionSR}\\
       &+D_{3}\hat{\eta}_{3}e^{ik_{S}^{+}x}+D_{4}\hat{\eta}_{4}e^{-ik_{S}^{-}x},\nonumber
    \end{align}
    where $\hat{\eta}_{1}=[1,0,0,R_{1}e^{i\phi/2}]^{T}$, $\hat{\eta}_{2}=[1,0,0,R_{2}e^{i\phi/2}]^{T}$, $\hat{\eta}_{3}=[0,1,-R_{1}e^{i\phi/2},0]^{T}$, and $\hat{\eta}_{4}=[0,1,-R_{2}e^{i\phi/2},0]^{T}$.

    The wave function in the F$_{1}$ layer is
    \begin{align}
       &\psi_{F_{1}}(x)=(M_{1}e^{ik_{1}x}+M_{1}^{\prime}e^{-ik_{1}x})\hat{e}_{1}+(M_{2}e^{ik_{2}x}+M_{2}^{\prime}e^{-ik_{2}x})\hat{e}_{2}\nonumber\\
       &+(M_{3}e^{ik_{3}x}+M_{3}^{\prime}e^{-ik_{3}x})\hat{e}_{3}+(M_{4}e^{ik_{4}x}+M_{4}^{\prime}e^{-ik_{4}x})\hat{e}_{4},\label{HM_wave}
      \end{align}
    where $\hat{e}_{1}=(1\;0\;0\;0)^{T}$, $\hat{e}_{2}=(0\;1\;0\;0)^{T}$, $\hat{e}_{3}=(0\;0\;1\;0)^{T}$, and $\hat{e}_{4}=(0\;0\;0\;1)^{T}$ are basis wave functions in the ferromagnetic region, and $k_{1(2)}=k_{F}\sqrt{1+(\epsilon \pm{h_{1}})/E_{F}-\left( k_{\parallel}/k_{F}\right) ^{2}}$ and $k_{3(4)}=k_{F}\sqrt{1-(\epsilon\mp{h_{1}})/{E_{F}}-\left( k_{\parallel}/{k_{F}}\right)^{2}}$ are the longitudinal components of the wave vectors for the quasiparticles in the F$_{1}$ layer. The corresponding wave function $\psi_{F_{2}}(x)$ in the F$_{2}$ layer can be obtained from Eq.~(\ref{HM_wave}) by replacement $h_{1}\rightarrow{h_{2}}$. It is worthy to note that the parallel component $k_{\parallel}$ is conserved in transport processes of the quasiparticles.

    The wave functions [$\psi_{L}^{S}(x)$, $\psi_{F1}(x)$, $\psi_{F2}(x)$, and $\psi_{R}^{S}(x)$] and their first derivatives should satisfy the boundary conditions at the S/F$_{1}$, F$_{1}$/F$_{2}$, and F$_{2}$/S interfaces,
    \begin{align}
         & \psi_{L}^{S}(-d_{1})=\psi_{F1}(-d_{1}),\nonumber\\
         & \frac{\partial\psi_{L}^{S}}{\partial{x}}\left\vert_{x=-d_{1}}\right.
          =\frac{\partial\psi_{F1}}{\partial{x}}\left\vert_{x=-d_{1}}\right.,\label{condition1}\\
         & \psi_{F1}(0)=\psi_{F2}(0),\nonumber\\
         & \frac{d\psi_{F_{2}}}{dx}\left\vert_{x=0^{+}}\right.-\frac{d\psi_{F_{1}}}{dx}\left\vert_{x=0^{-}}\right.=k_{F}
         \begin{pmatrix}
            \hat{W} & 0\\
            0 & \hat{W}^{*}
         \end{pmatrix}
         \psi(0),\label{boundary2}\\
         & \psi_{F2}(d_{2})=\psi_{R}^{S}(d_{2}),\nonumber\\
         & \frac{\partial\psi_{F2}}{\partial{x}}\left\vert_{x=d_{2}}\right.
         =\frac{\partial\psi_{R}^{S}}{\partial{x}}\left\vert_{x=d_{2}}\right.,\label{condition3}
    \end{align}
     where
    \begin{equation}
        \hat{W}=
        \begin{pmatrix}
            Z-P_{z} & -(P_{x}-iP_{y})\\
            -(P_{x}+iP_{y}) & Z+P_{z}
        \end{pmatrix}.
    \end{equation}
    We define the dimensionless spin-independent parameter $Z=2mV_{0}/(\hbar^{2}k_{F})$ measuring the strength of the potential barrier and the dimensionless spin-dependent parameter $P_{x,y,z}=2m\rho_{x,y,z}/(\hbar^{2}k_{F})$ describing the spin-active barrier at the F$_{1}$/F$_{2}$ interface. For simplicity, we just consider the effect of $y$-component $P_{y}$ and ignore the role of $x$- and $y$-components ($P_{x}$ and $P_{z}$).

    From these boundary conditions we can set up 24 linear equations in the following form:
    \begin{equation}
       \hat{A}X=\hat{B},\label{linearEq}%
    \end{equation}
    where $X$ contains 24 scattering coefficients and $\hat{A}$ is a $24\times24$ matrix. The solution of the characteristic equation
    \begin{equation}
       \det\hat{A}=0\label{characteristicEq}%
    \end{equation}
    allows one to identify two Andreev bound-state solutions for energies $E_{A\omega}$ ($\omega$=1, 2). Below we will consider the case of the short Josephson junction with a thickness much smaller than the superconducting coherence length $\xi$. In such a case the main contribution to the Josephson current is provided by the Andreev bound states (see, e.g.,~\cite{PFBagwell, CWJBeenakker}). In a one-dimensional (1D) SF$_{1}$F$_{2}$S junction, the Josephson current can be calculated by the general formula
    \begin{equation}
       I^{1d}(\phi)=\frac{2e}{\hbar}\frac{\partial\Omega}{\partial\phi},\label{current}
    \end{equation}
    where $\Omega$ is the phase-dependent thermodynamic potential. This potential can be obtained from the excitation spectrum by using the formula~\cite{JBardeen,JCayssol}
    \begin{equation}
        \Omega=-2T\sum_{\omega}\ln\left[ 2\cosh\frac{E_{A\omega}(\phi)}{2T}\right],\label{potential}
    \end{equation}
    where $\Delta$, $h_{1}$, $h_{2}$, $Z$ and $P_{y}$ are assumed to be the equilibrium values, which minimize the free energy of the SF$_{1}$F$_{2}$S structure and depend on microscopic parameters~\cite{Buzdin-AdvPhys85}. The summation in (\ref{potential}) is taken over all positive Andreev energies [$0<E_{A\omega}(\phi)<\Delta$]. For each value of $\phi$, we solve Eq.~(\ref{characteristicEq}) numerically to obtain the two spin-polarized Andreev levels. Since the Andreev energy spectra are doubled as they include the Bogoliubov redundancy, and only half of the energy states should be taken into account, we can find the 1D Josephson current via Eqs.~(\ref{current}) and (\ref{potential}).

    In a three-dimensional (3D) case, the Josephson current can be expressed as
    \begin{align}
        I^{3d}(\phi)&=\frac{S}{4\pi^{2}}\frac{2e\Delta}{\hbar}\int^{k_{F}}_{0}{I}^{1d}(k_{\parallel})2\pi{k_{\parallel}}dk_{\parallel}\nonumber\\
        &=\frac{4\pi\Delta}{eR_{N}}\int^{1}_{0}{I}^{1d}(\tilde{k}_{\parallel}){\tilde{k}_{\parallel}}d\tilde{k}_{\parallel},\label{3Dcurrent}
    \end{align}
    where $R^{-1}_{N}=e^{2}k^{2}_{F}S/(4\pi^{2}\hbar)$ is the Sharvin resistance and $\tilde{k}_{\parallel}=k_{\parallel}/k_{F}$ is the normalized wave vector. The 3D critical current can be derived from $I_{c}^{3d}=max_{\phi}|I^{3d}(\phi)|$.

    \section{Results and discussions}

    \begin{figure}
       \centering
       \includegraphics[width=3.55in]{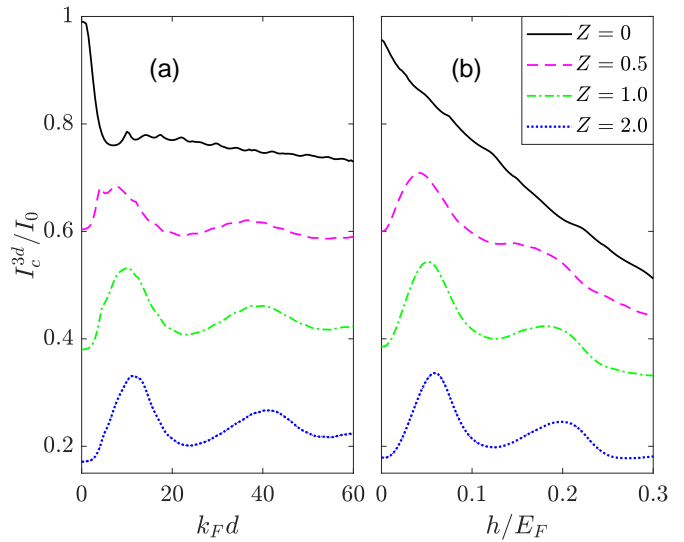} 
       \caption{The dependence of the 3D critical current $I_{c}^{3d}$ on the ferromagnetic thickness $k_Fd$ for the exchange field $h/E_F=0.1$ (a) and on the exchange field $h/E_F$ for the ferromagnetic thickness $k_Fd=20$ (b) when the potential barrier $Z$ takes several different values. Here we consider an antiparallel orientation of the exchange fields.}
       \label{Fig2}
    \end{figure}

     \begin{figure}[ptb]
     \centering
     \includegraphics[width=3.55in]{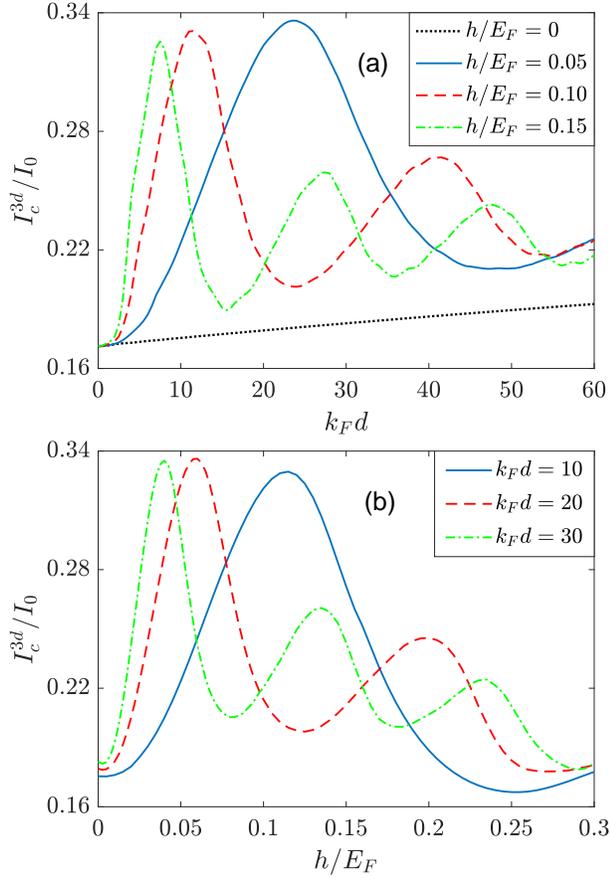}
     \caption{The dependence of the 3D critical current $I_{c}^{3d}$ on the ferromagnetic thickness $k_{F}d$ (a) and on the exchange field $h/E_{F}$ (b) in the case of an antiparallel orientation. Here the potential barrier is $Z=2$.}
     \label{Fig3}
     \end{figure}

     In our calculations we use the superconducting gap $\Delta$ as a unit of energy and take the Fermi energy $E_{F}=1000\Delta$. All lengths and the exchange field strengths are measured in units of the inverse Fermi wave vector $k_{F}$ and the Fermi energy $E_{F}$, respectively. Note that the approximation of the short Josephson junction ($k_Fd_1,k_Fd_2\ll1000$) is fully satisfied in the presented calculations. The normalized unit of current is $I_{0}=2\pi\Delta/(eR_{N})$ in the 3D case.

     We study the SF$_{1}$F$_{2}$S structures with a potential barrier $Z$ or a spin-active barrier $P_y$ at the F$_{1}$/F$_{2}$ interface. We present the results for $h_{1}=h_{2}=h$ for parallel exchange field and $h_{1}=-h_{2}=h$ for antiparallel exchange field, and define the ferromgnetic thickness $d_{1}=d_{2}=d$.

      \begin{figure}
         \centering
         \includegraphics[width=3.55in]{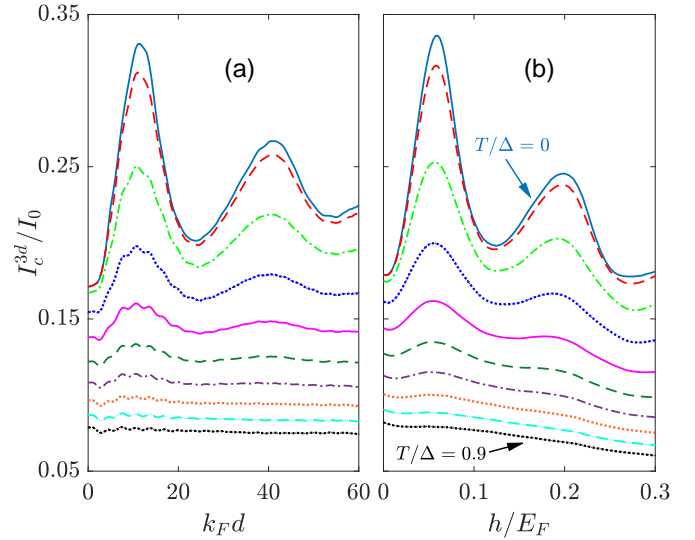} 
         \caption{The dependence of the 3D critical current $I_{c}^{3d}$ on the ferromagnetic thickness $k_Fd$ for the exchange field $h/E_F=0.1$ (a) and on the exchange field $h/E_F$ for the ferromagnetic thickness $k_Fd=20$ (b) in the case of an antiparallel orientation. Here the temperature $T/\Delta$ varies from 0 to 0.9 with a step 0.1, which corresponds to the curves from top to bottom. The potential barrier is $Z=2$.}
         \label{Fig4}
      \end{figure}

      We draw in Fig.~\ref{Fig2} the dependence of the critical current $I^{3d}_{c}$ on the ferromagnetic thickness $k_Fd$ and the exchange field $h/E_F$ for an antiparallel alignment of the magnetic moment $h_1=-h_2=h$ when the potential barrier $Z$ takes several different values. It is shown that the critical current decreases monotonically with the increasing exchange field for the transparent F$_1$/F$_2$ interface $Z=0$, while it reveals the oscillating behavior for $Z>0$. By increasing $Z$, the amplitude of the critical current decreases as a whole, but the oscillation behavior still remains. The critical current shows the same characteristic if one increases the ferromagnetic thickness $k_Fd$. These features indicate that the oscillation of the critical current originates from the resonant tunneling of the Cooper pairs occurring between the F$_1$ and F$_2$ layers. In fact the spin-dependent wave vector of the pairing electrons will change when the Cooper pairs pass through the F$_1$ and F$_2$ layer, and therefore the phase evolution of the Cooper pairs leads to the resonances occurring in F$_1$ and F$_2$. Therefore, the oscillation period depends on the exchange field and/or thickness of the ferromagnets, not on the properties of the central insulating barrier.

      \begin{figure}[ptb]
      \centering
      \includegraphics[width=3.55in]{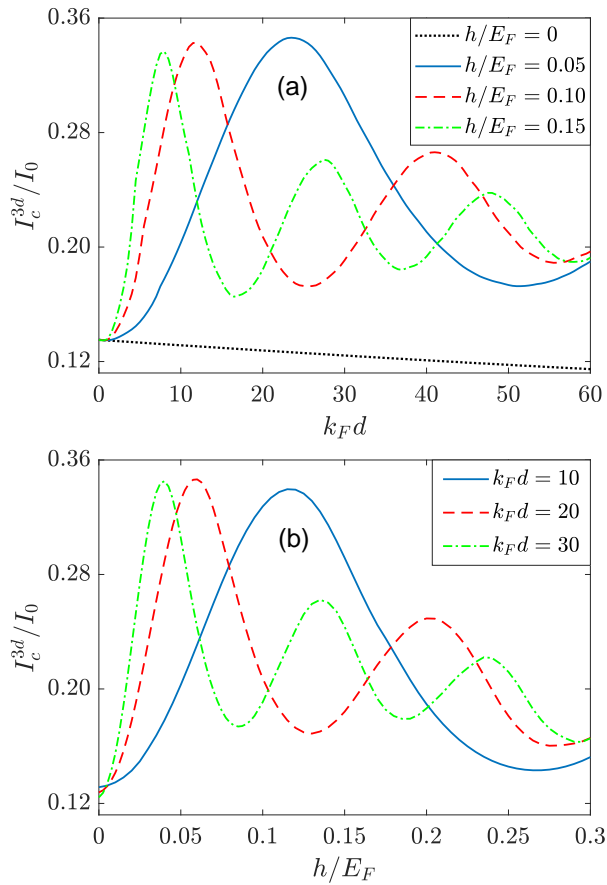}
      \caption{The dependence of the 3D critical current $I_{c}^{3d}$ on the ferromagnetic thickness $k_{F}d$ (a) and on the exchange field $h/E_{F}$ (b) in the case of a parallel orientation. Here the spin-active barrier is taken as $P_{y}=2$.}%
      \label{Fig5}
      \end{figure}

      If one changes the exchange field $h/E_{F}$ and thickness $k_{F}d$ of the ferromagnetic layers, the oscillation period of the critical current changes accordingly. The calculation results are illustrated in Fig.~\ref{Fig3}. The observed oscillations remind us of the oscillations observed previously in~\cite{CitD} for the 1D model of the junction with noncollinear magnetization and attributed to the geometrical resonances. The interesting consequence of the presence of barrier is the counterintuitive increase of the critical current with increasing exchange field [up to $h_{\rm {max}}/E_F\sim0.12$ when $k_Fd=10$, see Fig.~\ref{Fig3}(b)] or ferromagnetic layer thickness [up to $k_Fd_{\rm {max}}\sim24$ when $h/E_F=0.05$, see Fig.~\ref{Fig3}(a)]. Note that the similar increase of the current with the exchange field was obtained in the models of S/F tunnel structures~\cite{FSBerg,VNKrivo,Elena}. The key difference between our results and Refs.~\cite{FSBerg,VNKrivo,Elena} is that the initial increase was not followed by the oscillatory behavior of the critical current with exchange field and/or ferromagnetic layer thickness. We find also that the growth range of the critical current with the exchange field strongly depends on the ferromagnetic layer thicknesses. Moreover, note that in presence of a potential barrier the critical current slightly increases with normal-metal thickness when both ferromagnets become the normal-metal ($h/E_F=0$) [see Fig.~\ref{Fig3}(a)]. This circumstance reflects the presence of some resonance effects in this case too. It should be noted that the oscillatory effect mentioned above is revealed only at low temperatures. The dependence of the critical current on the temperature is illustrated in Fig.~\ref{Fig4}. We see that the oscillations of the critical current $I^{3d}_c$ will decrease as the temperature $T/\Delta$ increases, which should be related to the smearing of the resonance tunneling from the lowest Andreev levels. When $T/\Delta$ reaches 0.9, the oscillation completely disappears and $I^{3d}_c$ decreases monotonously.

      \begin{figure}[ptb]
         \centering
         \includegraphics[width=3.55in]{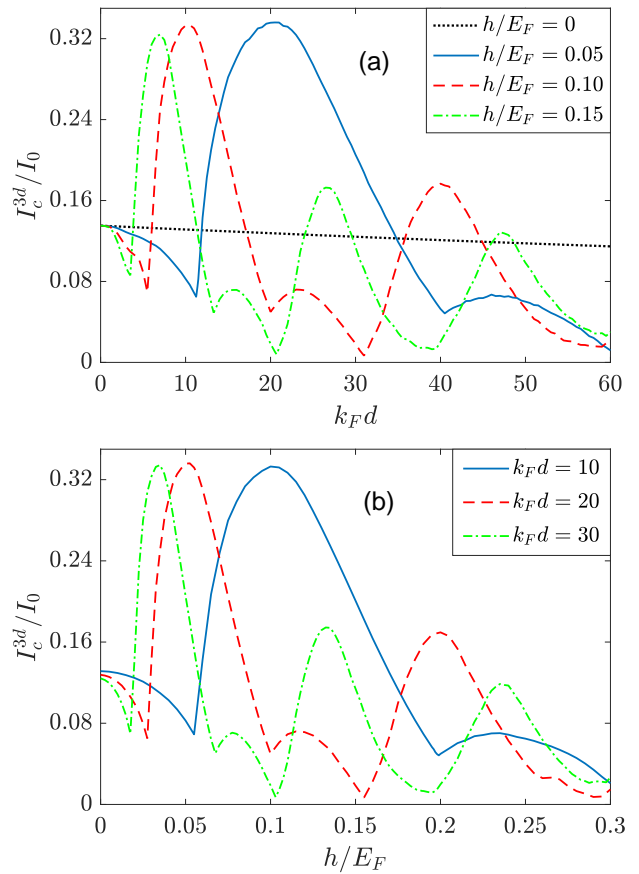}
         \caption{The dependence of the 3D critical current $I_{c}^{3d}$ on the ferromagnetic thickness $k_{F}d$ (a) and on the exchange field $h/E_{F}$ (b) in the case of an antiparallel orientation. Here the spin-active barrier is taken as $P_{y}=2$. The dips at the curves signal the transitions between 0 and $\pi$ states.}
         \label{Fig6}
      \end{figure}

      In Fig.~\ref{Fig5}, we show the variation characteristics of the critical current in SF$_{1}$F$_{2}$S structure with the magnetic moment in F$_1$ and F$_2$ being parallel and with a spin-active barrier at the F$_1$/F$_2$ interface. It is found that the critical current also displays an oscillating behavior. The behavior is similar to the cases in which the magnetic moments are antiparallel and there is a potential barrier at the F$_1$/F$_2$ interface. So we can say that the spin-active barrier at the F$_1$/F$_2$ interface can play two roles: (i) It creates a spin-flip effect to flip the spin of the conduction electrons crossing the F$_1$/F$_2$ interface. The two ferromagnets have the same energy band because of the parallel polarized direction of the magnetic moments. In such a case, spin-$\uparrow$ ($\downarrow$) electrons will be transformed into spin-$\downarrow$ ($\uparrow$) electrons when they pass from the F$_1$ layer into the F$_2$ layer. The same electron will occupy the opposite spin band in the F$_1$ and F$_2$ layers. This situation is similar to the antiparallel ferromagnets without the spin-flip in the central interface. (ii) It acts as a potential barrier, which hinders electron tunneling and reduces the transmission of the F$_1$/F$_2$ interface. Therefore, we can still see the oscillating phenomenon of the critical current in this structure.

      \begin{figure}[ptb]
         \centering
         \includegraphics[width=3.55in]{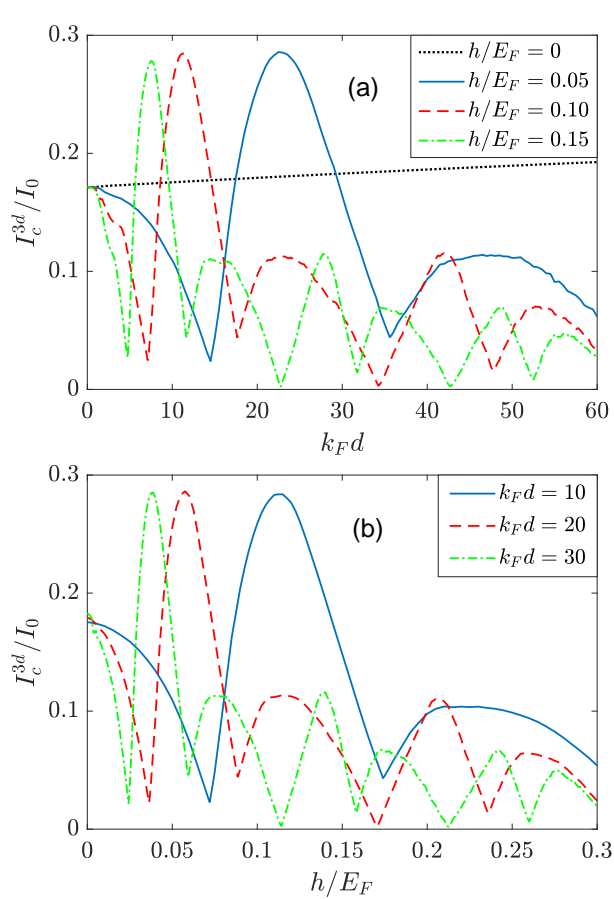}
         \caption{The dependence of the 3D critical current $I_{c}^{3d}$ on the ferromagnetic thickness $k_{F}d$ (a) and on the exchange field $h/E_{F}$ (b) in the case of a parallel orientation. Here the potential barrier is taken as $Z=2$. The dips at the curves signal the transitions between 0 and $\pi$ states.}
         \label{Fig7}
      \end{figure}

      Similarly, the above two roles caused by the spin-active barrier can also present in the antiparallel SF$_{1}$F$_{2}$S junction. If one only considers the role of the spin-flip effect, the antiparallel SF$_{1}$F$_{2}$S junction with a central spin-flip is equivalent to a homogenous SFS junction. In this case, the 0-$\pi$ transition will resume. For example, at $h/E_F=0.05$ the inversion of the current sign takes place at $k_Fd\approx10$ and $k_Fd\approx40$ (see Fig.~\ref{Fig6}(a) and Fig. 1 in Supplemental Material~\cite{SPMaterial}). In other words, the junction is in $\pi$ state for $k_Fd\prec10$ and $k_Fd\succ40$, as well as it will become 0 state in the region $10\prec{k_Fd}\prec40$. Moreover, the insulating property of the spin-active barrier causes a resonant tunneling of electrons. This results in the largest peaks that appear periodically in the current $I_{c}^{3d}$. For example, if one looks at the curve for $h/E_F=0.10$ in Fig~\ref{Fig6}(a), the resonance produces the peaks at $k_Fd\approx10$ and $k_Fd\approx40$, which appear in similar positions in Fig.~\ref{Fig3}(a).

      \begin{figure}[ptb]
         \centering
         \includegraphics[width=3.55in]{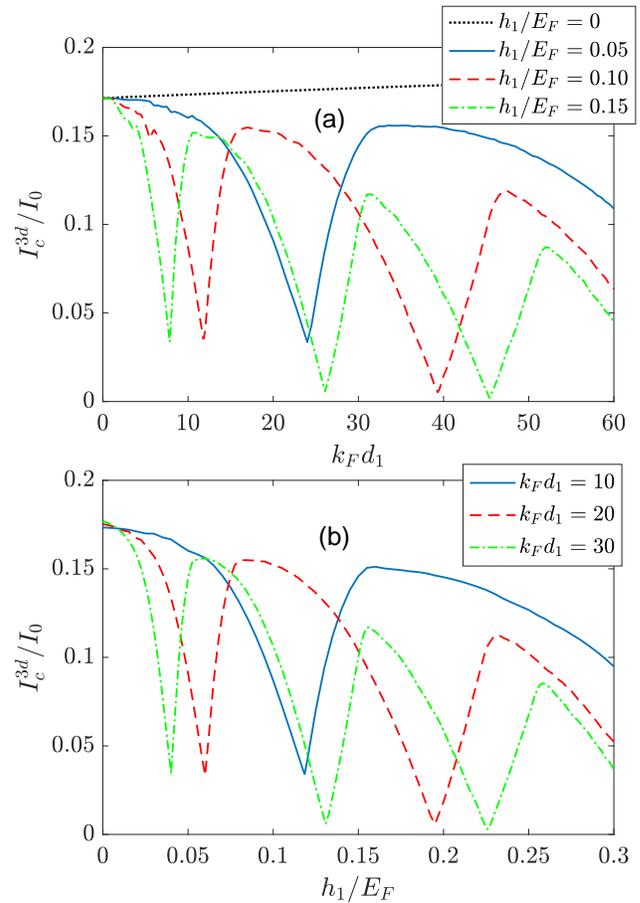}
         \caption{The dependence of the 3D critical current $I_{c}^{3d}$ on the ferromagnetic thickness $k_{F}d_1$ (a) and on the exchange field $h_1/E_{F}$ (b) for the SF$_1$S configuration ($d_2=0$) with the potential barrier $Z=2$ at the right F$_1$/S interface. The dips at the curves signal the transitions between 0 and $\pi$ states.}
         \label{Fig8}
      \end{figure}

      \begin{figure}[ptb]
         \centering
         \includegraphics[width=3.55in]{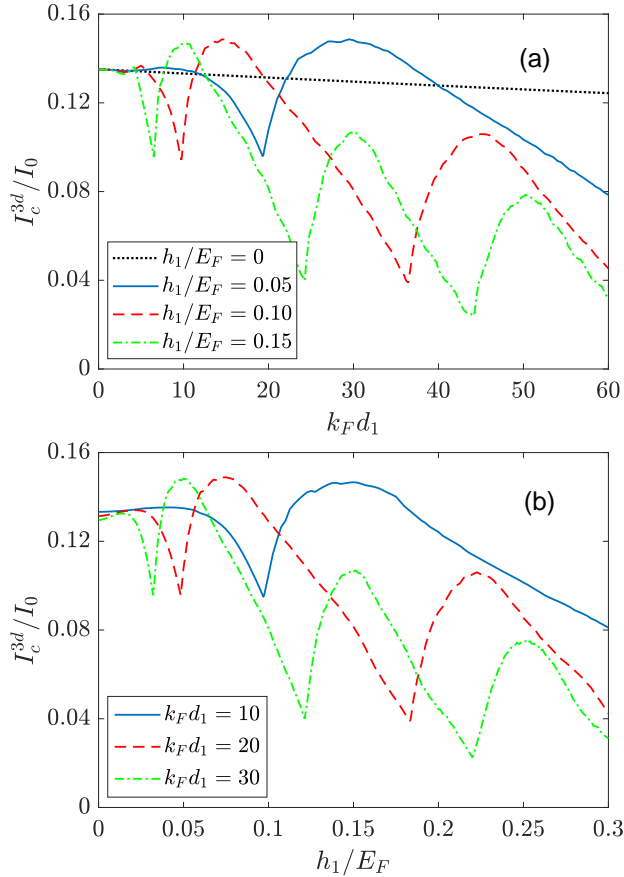}
         \caption{The dependence of the 3D critical current $I_{c}^{3d}$ on the ferromagnetic thickness $k_{F}d_1$ (a) and on the exchange field $h_1/E_{F}$ (b) for the SF$_1$S configuration ($d_2=0$) with the spin-active barrier $P_y=2$ at the right F$_1$/S interface. The dips at the curves signal the transitions between 0 and $\pi$ states.}
         \label{Fig9}
      \end{figure}

      To further demonstrate the coexistence of resonant tunneling and 0-$\pi$ transition, we calculated the current in the parallel SF$_1$F$_2$S junction with a central potential barrier. It is known that, in a uniform SFS junction, the critical current decays with increasing ferromagnetic thickness (or exchange field) and also reveals oscillations caused by the 0-$\pi$ transition. If the potential barrier is introduced at the center of the ferromagnet, the amplitude of the critical current will be suppressed overall because the potential barrier reduces the transmission of the conduction electrons. Meanwhile, the resonant tunneling of the conduction electrons between F$_1$ and F$_2$ layers induces the periodic peaks in the critical current. As a result, the critical current shows singular features in Fig.~\ref{Fig7} and Fig. 2 of the Supplemental Material~\cite{SPMaterial}.

      Finally, in order to illustrate the previous conjecture regarding resonant tunneling, we discuss the current in the SF$_1$S junction ($d_2=0$) with the potential or spin-active barriers at the F$_1$/S interface. As shown in Figs.~\ref{Fig8} and \ref{Fig9}, the critical current displays a damped oscillation with increasing thickness $k_Fd_1$ and/or exchange field $h_1/E_F$. These current oscillations can be attributed to the 0-$\pi$ transition (see Figs. 3 and 4 in the Supplemental Material~\cite{SPMaterial}) but not to the periodic peaks induced by the resonant tunneling between the F$_1$ and F$_2$ layers, because the resonant tunneling cannot exist in these structures. In addition, we find that the critical current at the transition between the 0 and $\pi$ states is close to zero in the SF$_1$S junction with the potential barrier $Z=2$ at the F$_1$/S interface, when the thickness $k_Fd_1$ and/or exchange field $h_1/E_F$ take larger values. However, this current is much larger when the F$_1$/S interface has a spin-active barrier $P_y=2$ (see Fig.~\ref{Fig9}). This may be related to the important contribution from the second harmonic current in the presence of spin-active interface structure~\cite{FSBerg,Luka,Caroline,ASMel}.

     \section{Conclusion}

     On the basis of the exact numerical solution of the Bogoliubov-de Gennes equations, we have studied the Josephson current in the SF$_1$F$_2$S junctions containing a potential or spin-active barrier at F$_1$/F$_2$ interface. We show that at low temperature the potential barrier may result in large oscillations of the critical current as a function of the ferromagnetic layer thickness and exchange field even for the antiparallel orientation of the magnetic moment in the F$_1$ and F$_2$ layers. Such behavior is related to the interference effects of the electrons wave functions and may be considered as some form of the geometrical resonance phenomena. Specifically, comparing to the normal-metal junction ($h=0$ in our model), the exchange field ($h>0$) can enhance the critical current for the antiparallel configuration. In contrast, the spin-active barrier in this antiparallel configuration leads to the 0-$\pi$ transitions, which is similar to the case of uniform SFS junction. The spin-active barrier in the parallel configuration can also cause the oscillations of the critical current. The obtained results may be useful for the interpretation of the experimental data on the Josephson junctions with composite ferromagnetic barrier.

     \section*{Acknowledgments}

     The authors thank A. Melnikov and S. Mironov for useful discussions and suggestions. This work was supported by French ANR projects SUPERTRONICS and OPTOFLUXONICS, EU Network COST CA16218 (NANOCOHYBRI), ANR-DFG grant ``Fermi-NESt'', and ERC 647100 ``SUSPINTRONICS''. H. Meng was supported by the National Natural Science Foundation of China (Grant No.11604195 and No.11447112) and the Youth Hundred Talents Programme of Shaanxi Province.


\begin{thebibliography}{99}

\bibitem {Golubov}A. A. Golubov, M. Yu. Kupriyanov, and E. Ilichev, The
current-phase relation in Josephson junctions, Rev. Mod. Phys. \textbf{76},
411 (2004).

\bibitem {Buzdin}A. I. Buzdin, Proximity effects in superconductor-ferromagnet
heterostructures, Rev. Mod. Phys. \textbf{77}, 935 (2005).

\bibitem {Bergeret}F. S. Bergeret, A. F. Volkov, and K. B. Efetov, Odd triplet
superconductivity and related phenomena in superconductor-ferromagnet
structures, Rev. Mod. Phys. \textbf{77}, 1321-1373 (2005).

\bibitem {Linder}J. Linder and J. W. A. Robinson, Superconducting spintronics,
Nat. Phys. \textbf{11}, 307 (2015).

\bibitem {Eschrig}M. Eschrig, Spin-polarized supercurrents for spintronics: a
review of current progress, Rep. Prog. Phys. \textbf{78}, 104501 (2015).

\bibitem {Eilenberger}G. Eilenberger, Transformation of Gorkov's equation for
type II superconductors into transport-like equations, Z. Phys. \textbf{214},
195-213 (1968).

\bibitem {Usadel}K. D. Usadel, Generalized Diffusion Equation for
Superconducting Alloys, Phys. Rev. Lett. \textbf{25}, 507 (1970).

\bibitem {Reeg}C. R. Reeg and D. L. Maslov, Proximity-induced triplet
superconductivity in Rashba materials, Phys. Rev. B \textbf{92}, 134512 (2015).

\bibitem {Silaev}M. A. Silaev, I. V. Tokatly, and F. S. Bergeret, Anomalous
current in diffusive ferromagnetic Josephson junctions, Phys. Rev. B
\textbf{95}, 184508 (2017).

\bibitem {HaoBuzdin}Hao Meng, A. V. Samokhvalov, and A. I. Buzdin, Nonuniform
superconductivity and Josephson effect in a conical ferromagnet, Phys. Rev. B
\textbf{99}, 024503 (2019).



\bibitem {CVisani} C. Visani, Z. Sefrioui, J. Tornos, C. Leon, J. Briatico, M. Bibes, A. Barth\'{e}l\'{e}my, J. Santamar\'{\i}a, and Javier E. Villegas,Equal-spin Andreev reflection and long-range coherent transport in high-temperature superconductor/half-metallic ferromagnet junctions, Nature Physics 8, 539-543 (2012).

\bibitem {CVFC} C. Visani, F. Cuellar, A. P\'{e}rez-Mu\~{n}oz, Z. Sefrioui, C. Le\'{o}n, J. Santamar\'{\i}a, and Javier E. Villegas, Magnetic field influence on the proximity effect at YBa$_2$Cu$_3$O$_7$/La$_{2/3}$Ca$_{1/3}$MnO$_3$ superconductor/half-metal interfaces, Phys. Rev. B 92, 014519 (2015).

\bibitem {MJWA} M. Egilmez, J. W. A. Robinson, Judith L. MacManus-Driscoll, L. Chen, H. Wang and M. G. Blamire, Supercurrents in half-metallic ferromagnetic La$_{0.7}$Ca$_{0.3}$MnO$_3$, Europhys. Lett. 106, 37003 (2014).


\bibitem {PGdeGennes}P. G. de Gennes, Superconductivity of Metals and Alloys,
Benjamin, New York, 1966 (Chap.5).



  \bibitem {CitE} Z. Radovi\'{c}, N. Lazarides, and N. Flytzanis, Josephson effect in double-barrier superconductor-ferromagnet junctions, Phys. Rev. B \textbf{68}, 014501 (2003).

  \bibitem {CitC} Klaus Halterman and Oriol T. Valls, Layered ferromagnet-superconductor structures: The $\pi$ state and proximity effects, Phys. Rev. B \textbf{69}, 014517 (2004).

  \bibitem {CitB} Paul H. Barsic, Oriol T. Valls, and Klaus Halterman, Thermodynamics and phase diagrams of layered superconductor/ferromagnet nanostructures, Phys. Rev. B \textbf{75}, 104502 (2007).

  \bibitem {CitA} Klaus Halterman, Oriol T. Valls, and Chien-Te Wu, Charge and spin currents in ferromagnetic Josephson junctions, Phys. Rev. B \textbf{92}, 174516 (2015).

  \bibitem {CitD} Z. Pajovi\'{c}, M. Bo\v{z}ovi\'{c}, Z. Radovi\'{c}, J. Cayssol, and A. Buzdin, Josephson coupling through ferromagnetic heterojunctions with noncollinear magnetizations, Phys. Rev. B \textbf{74}, 184509 (2006).


  \bibitem {Hman2} Klaus Halterman and Mohammad Alidoust, Half-metallic superconducting triplet spin valve, Phys. Rev. B \textbf{94}, 064503 (2016).
  \bibitem {Hman3} Klaus Halterman and Mohammad Alidoust, Josephson currents and spin-transfer torques in ballistic SFSFS nanojunctions, Supercond. Sci. Technol. \textbf{29}, 055007 (2016).
  \bibitem {Hman4} Mohammad Alidoust and Klaus Halterman, Half-metallic superconducting triplet spin multivalves, Phys. Rev. B \textbf{97}, 064517 (2018).
  \bibitem {Hman5} Klaus Halterman and Mohammad Alidoust, Induced energy gap in finite-sized superconductor/ferromagnet hybrids, Phys. Rev. B \textbf{98}, 134510 (2018).
  \bibitem {Hman6} Chien-Te Wu and Klaus Halterman, Spin transport in half-metallic ferromagnet-superconductor junctions, Phys. Rev. B \textbf{98}, 054518 (2018).


  \bibitem {CBell} C. Bell, G. Burnell, C. W. Leung, E. J. Tarte,D.-J. Kang, and M. G. Blamire, Controllable Josephson current through a pseudospin-valve structure, Appl. Phys. Lett. \textbf{84}, 1153-1155 (2004).

  \bibitem {JWAR} J. W. A. Robinson, Gabor B. Halasz, A. I. Buzdin, and M. G. Blamire, Enhanced Supercurrents in Josephson Junctions Containing Nonparallel Ferromagnetic Domains, Phys. Rev. Lett. \textbf{104}, 207001 (2010).

  \bibitem {BBa} B. Baek, W. H. Rippard, S. P. Benz, S. E. Russek, and P. D. Dresselhaus, Hybrid superconducting-magnetic memory device using competing order parameters, Nat. Commun. \textbf{5}, 3888 (2014).

  \bibitem {MAEQ} M. A. E. Qader, R. K. Singh, S. N. Galvi, L. Yu, J. M. Rowell, and N. Newman, Switching at small magnetic fields in Josephson junctions fabricated with ferromagnetic barrier layers, Appl. Phys. Lett. \textbf{104}, 022602 (2014).

  \bibitem {BBWHR} B. Baek, W. H. Rippard, M. R. Pufall, S. P. Benz, S. E. Russek, H. Rogalla, and P. D. Dresselhaus, Spin-Transfer Torque Switching in Nanopillar Superconducting-Magnetic Hybrid Josephson Junctions, Phys. Rev. Applied \textbf{3}, 011001 (2015).

  \bibitem {ECGing} E. C. Gingrich, Bethany M. Niedzielski, Joseph A. Glick, Yixing Wang, D. L. Miller, Reza Loloee, W. P. Pratt Jr, and Norman O. Birge, Controllable 0-$\pi$ Josephson junctions containing a ferromagnetic spin valve, Nat. Phys. \textbf{12}, 564 (2016).

  \bibitem {BMNied} Bethany M. Niedzielski, T. J. Bertus, Joseph A. Glick, R. Loloee, W. P. Pratt, Jr., and Norman O. Birge, Spin-valve Josephson junctions for cryogenic memory, Phys. Rev. B \textbf{97}, 024517 (2018).


   \bibitem {CitF} A. V. Samokhvalov, R. I. Shekhter, and A. I. Buzdin, Stimulation of a singlet superconductivity in SFS weak links by spin-exchange scattering of Cooper pairs, Sci. Rep. \textbf{4}, 5671 (2014).

   \bibitem {CitG} Iryna Kulagina and Jacob Linder, Spin supercurrent, magnetization dynamics, and $\varphi$-state in spin-textured Josephson junctions, Phys. Rev. B \textbf{90}, 054504 (2014).


   \bibitem {FSBerg} F. S. Bergeret, A. F. Volkov, and K. B. Efetov, Enhancement of the Josephson Current by an Exchange Field in Superconductor-Ferromagnet Structures, Phys. Rev. Lett. \textbf{86}, 3140 (2001).
   \bibitem {VNKrivo} V. N. Krivoruchko and E. A. Koshina, From inversion to enhancement of the dc Josephson current in S/F-I-F/S tunnel structures, Phys. Rev. B \textbf{64}, 172511 (2001).
   \bibitem {Elena} Elena Koshina and Vladimir Krivoruchko, Spin polarization and $\pi$-phase state of the Josephson contact: Critical current of mesoscopic SFIFS and SFIS junctions, Phys. Rev. B \textbf{63}, 224515 (2001).
   \bibitem {AAGolu} A. A. Golubov, M. Yu. Kupriyanov, and Ya. V. Fominov, Critical current in SFIFS junctions, Pis'ma Zh. Eksp. Teor. Fiz. \textbf{75}, 223 (2002) [JETP Lett. \textbf{75}, 190 (2002)].

   \bibitem {YaMB} Ya. M. Blanter and F. W. J. Hekking, Supercurrent in long SFFS junctions with antiparallel domain configuration, Phys. Rev. B \textbf{69}, 024525 (2004).

   \bibitem {BCrou} B. Crouzy, S. Tollis, and D. A. Ivanov, Josephson current in a superconductor-ferromagnet junction with two noncollinear magnetic domains, Phys. Rev. B \textbf{75}, 054503 (2007).


   \bibitem {PFBagwell} Philip F. Bagwell, Suppression of the Josephson current through a narrow, mesoscopic, semiconductor channel by a single impurity, Phys. Rev. B \textbf{46}, 12573 (1992).

   \bibitem {CWJBeenakker} C. W. J. Beenakker, Universal limit of critical-current fluctuations in mesoscopic Josephson junctions, Phys. Rev. Lett. \textbf{67}, 3836 (1991).




   \bibitem {JBardeen} J. Bardeen, R. K\"{u}mel, A. E. Jacobs, and L. Tewordt, Structure of Vortex Lines in Pure Superconductors, Phys. Rev. \textbf{187}, 556 (1969).

   \bibitem {JCayssol} J. Cayssol and G. Montambaux, Exchange-induced ordinary reflection in a single-channel superconductor-ferromagnet-superconductor junction, Phys. Rev. B \textbf{70}, 224520 (2004).

   \bibitem {Buzdin-AdvPhys85} L. N. Bulaevskii, A. I. Buzdin, M. L. Kuli\'{c}, and S. V. Panyukov, Coexistence of superconductivity and magnetism theoretical predictions and experimental results, Adv. Phys. \textbf{34}, 175-261 (1985).


   \bibitem {SPMaterial} See Supplemental Material for comparison between the absolute value of critical current $|I_c^{3d}|$ and the original critical current $I_c^{3d}$.

   \bibitem {Luka} Luka Trifunovic, Long-Range Superharmonic Josephson Current, Phys. Rev. Lett. \textbf{107}, 047001 (2011).

   \bibitem {Caroline} Caroline Richard, Manuel Houzet, and Julia S. Meyer, Superharmonic Long-Range Triplet Current in a Diffusive Josephson Junction, Phys. Rev. Lett. \textbf{110}, 217004 (2013).

   \bibitem {ASMel} A. S. Mel¡¯nikov, A. V. Samokhvalov, S. M. Kuznetsova, and A. I. Buzdin, Interference Phenomena and Long-Range Proximity Effect in Clean Superconductor-Ferromagnet Systems, Phys. Rev. Lett. \textbf{109}, 237006 (2012).

\end{thebibliography}
\end{document}


\title{Supplementary material for ``Josephson current through a ferromagnetic bilayer: beyond the quasiclassical approximation''}
\author{Hao Meng}
\affiliation{School of Physics and Telecommunication Engineering, Shaanxi University of Technology, Hanzhong 723001, China}
\affiliation{University Bordeaux, LOMA UMR-CNRS 5798, F-33405 Talence Cedex, France}
\author{Yajie Ren}
\affiliation{School of Physics and Telecommunication Engineering, Shaanxi University of Technology, Hanzhong 723001, China}
\author{Javier E. Villegas}
\affiliation{Unit\'{e} Mixte de Physique CNRS/Thales, Universit\'{e} Paris-Sud, Universit\'{e} Paris Saclay, 1 Avenue A. Fresnel, 91767 Palaiseau, France}
\author{A. I. Buzdin}
\email{alexandre.bouzdine@u-bordeaux.fr}
\affiliation{University Bordeaux, LOMA UMR-CNRS 5798, F-33405 Talence Cedex, France}
\affiliation{Sechenov First Moscow State Medical University, Moscow, 119991, Russia}
\date{\today}

    \maketitle

    In our article, the critical current is defined by formula $I_c^{3d}=max_{\phi}|I^{3d}(\phi)|$, so one cannot see the inversion of the current sign, which corresponds to the 0-$\pi$ transition. In order to clearly show the 0-$\pi$ transition, we plot the absolute value of critical current $|I_c^{3d}|$ and the original critical current $I_c^{3d}$ in the following figures.

    Figure~\ref{Figs1} shows the critical current in the antiparallel SF$_1$F$_2$S junction with the central spin-active barrier $P_y=2$. Meanwhile, in Fig.~\ref{Figs2} we plot the critical current in the parallel SF$_1$F$_2$S junction with the central potential barrier $Z=2$. From the right columns in the above two figures, we can see that the critical current $I_c^{3d}$ changes between negative and positive values. This feature demonstrates that the crossover between 0 and $\pi$ states takes place.

    In addition, we calculate the critical current in the SF$_1$S junction ($d_2=0$) with potential barrier $Z=0$ and spin-active barrier $P_y=2$ at F$_1$/S interface. Corresponding results are illustrated in Figs.~\ref{Figs3} and \ref{Figs4}, respectively. The 0-$\pi$ transition can also be seen in these two cases.

   \begin{figure*}[ptb]
   \centering
   \includegraphics[width=7.1in]{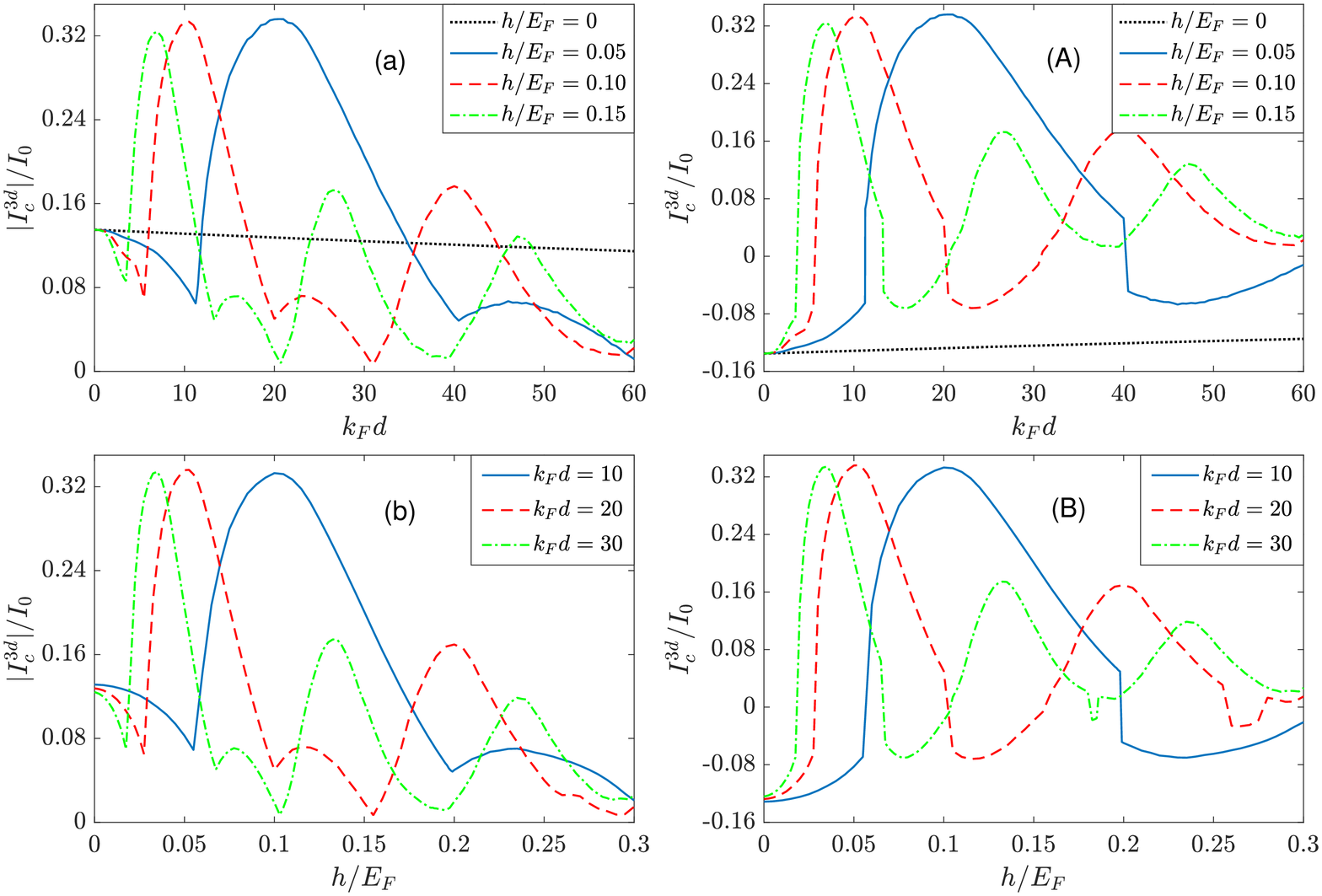}
    \caption{The dependence of the absolute value of critical current $|I_c^{3d}|$ (left column) and the original critical current $I_c^{3d}$ (right column) on the ferromagnetic thickness $k_Fd$ [(a) and (A)] and on the exchange field $h/E_F$ [(b) and (B)] for the antiparallel SF$_1$F$_2$S junction with the central spin-active barrier $P_y=2$. Figures in left column correspond to Fig.6 in our article.}
   \label{Figs1}
   \end{figure*}

   \begin{figure*}[ptb]
   \centering
   \includegraphics[width=7.1in]{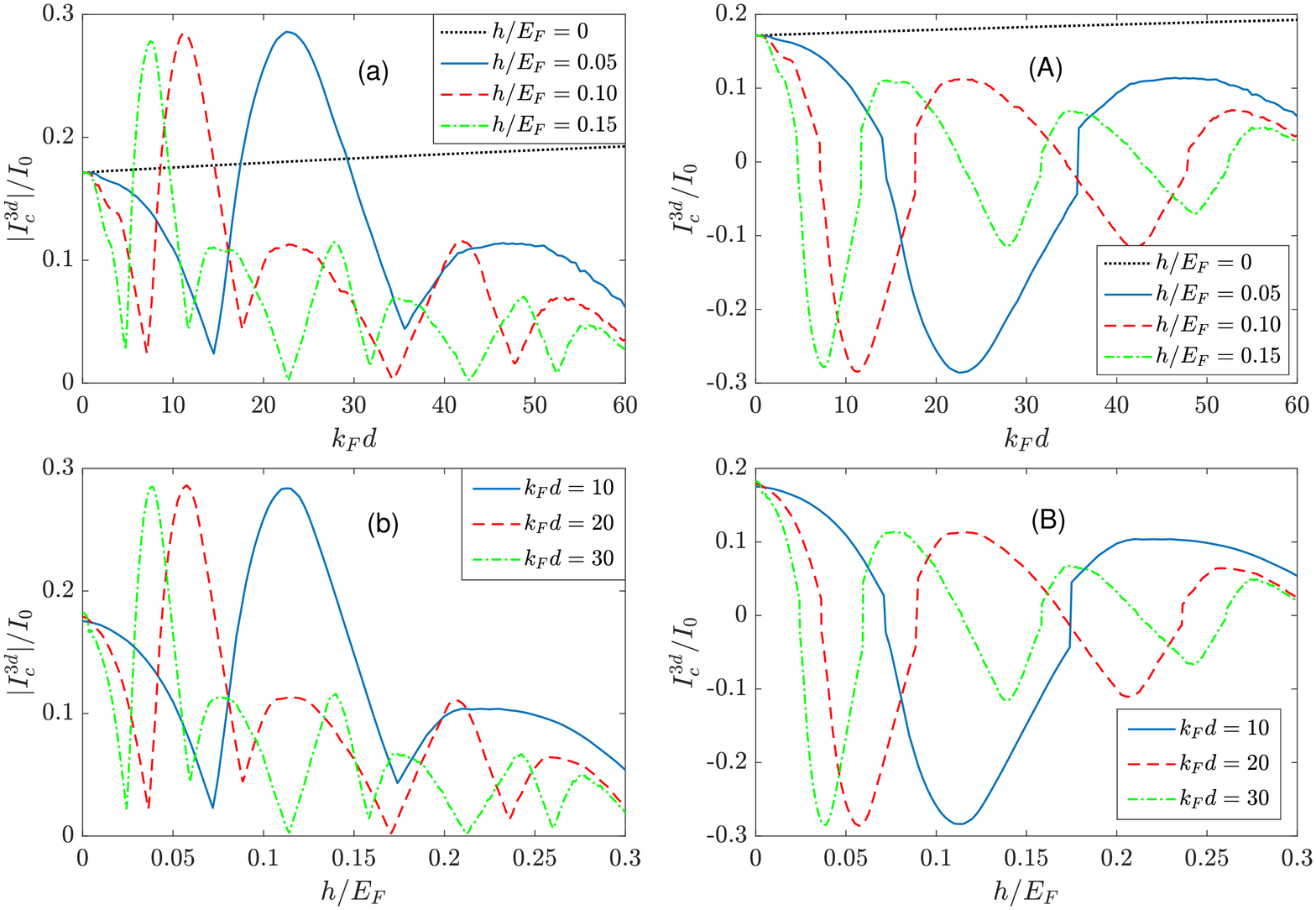}
    \caption{The dependence of the absolute value of critical current $|I_c^{3d}|$ (left column) and the original critical current $I_c^{3d}$ (right column) on the ferromagnetic thickness $k_Fd$ [(a) and (A)] and on the exchange field $h/E_F$ [(b) and (B)] for the parallel SF$_1$F$_2$S junction with the central potential barrier $Z=2$. Figures in left column correspond to Fig.7 in our article.}
   \label{Figs2}
   \end{figure*}

   \begin{figure*}[ptb]
   \centering
   \includegraphics[width=7.1in]{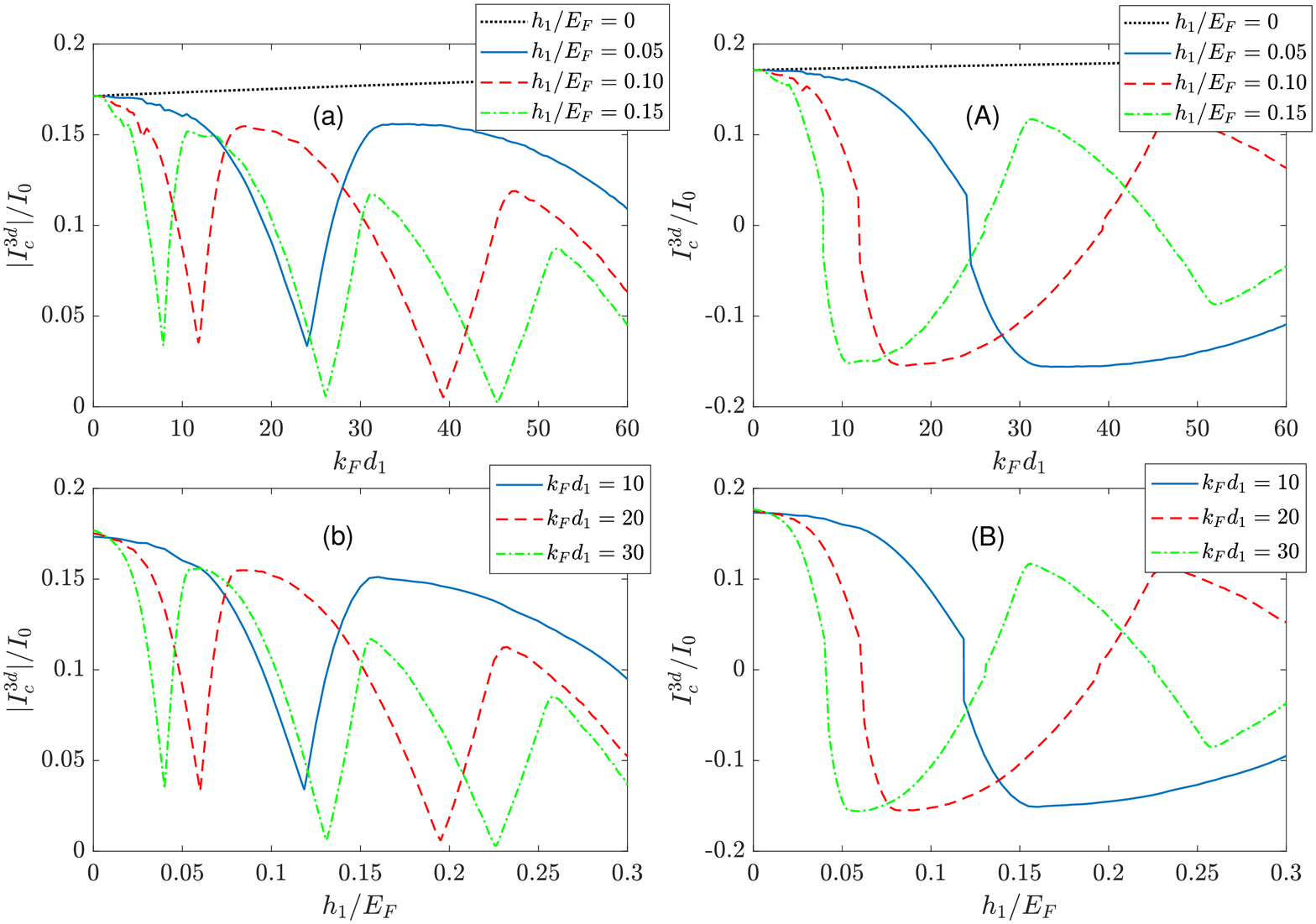}
    \caption{The dependence of the absolute value of critical current $|I_c^{3d}|$ (left column) and the original critical current $I_c^{3d}$ (right column) on the ferromagnetic thickness $k_Fd_1$ [(a) and (A)] and on the exchange field $h_1/E_F$ [(b) and (B)] for SF$_1$S junction with potential barrier $Z=2$ at F$_1$/S interface. Figures in left column correspond to Fig.8 in our article.}
   \label{Figs3}
   \end{figure*}

   \begin{figure*}[ptb]
   \centering
   \includegraphics[width=7.1in]{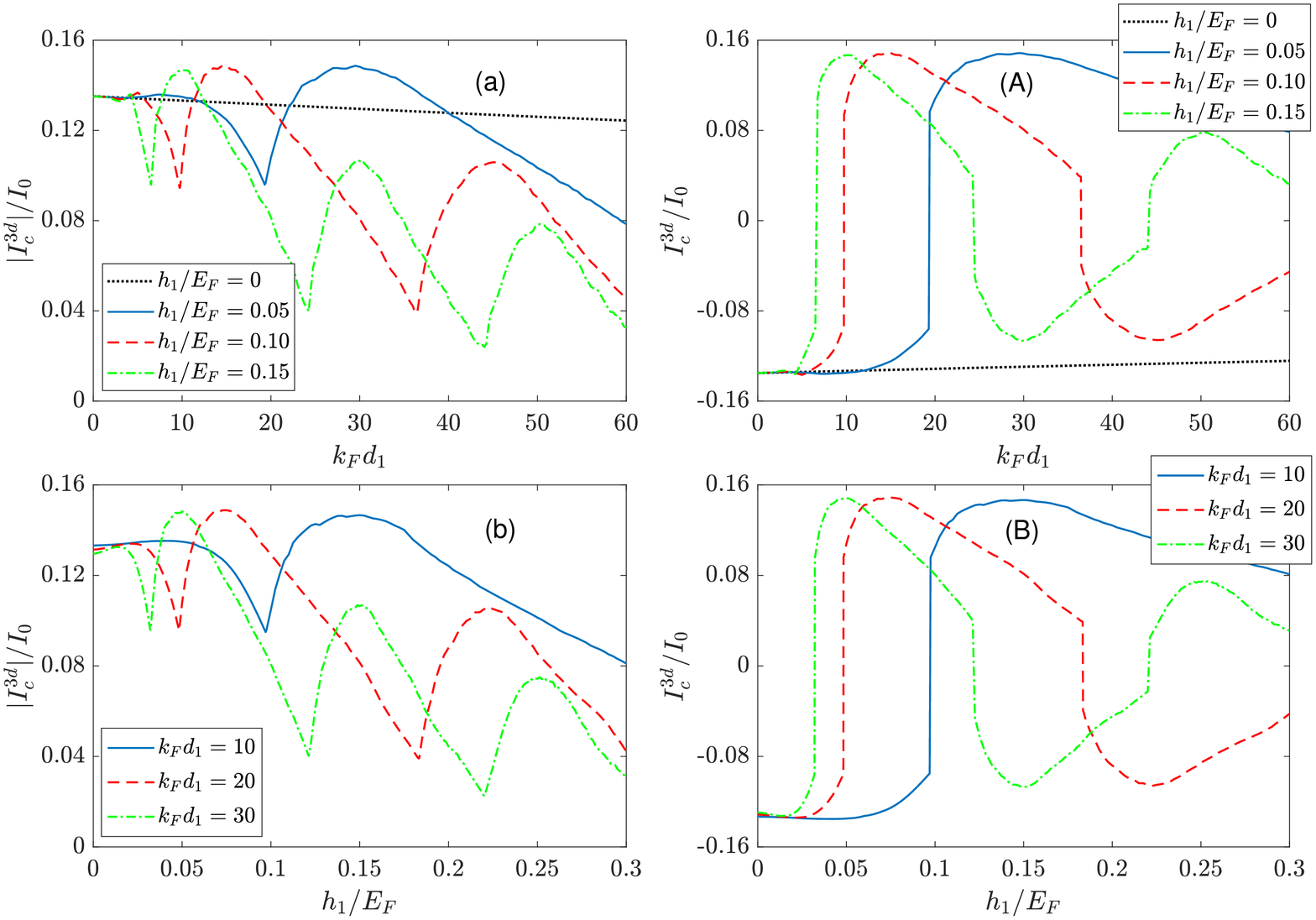}
    \caption{The dependence of the absolute value of critical current $|I_c^{3d}|$ (left column) and the original critical current $I_c^{3d}$ (right column) on the ferromagnetic thickness $k_Fd_1$ [(a) and (A)] and on the exchange field $h_1/E_F$ [(b) and (B)] for SF$_1$S junction with spin-active barrier $P_y=2$ at F$_1$/S interface. Figures in left column correspond to Fig.9 in our article.}
   \label{Figs4}
   \end{figure*}